# Information Theoretic I-MMSE generalize Time-Frequency Signal Processing Tools

Samah A. M. Ghanem

Senior Member, IEEE

***Abstract*—In this paper, we capitalize on information theoretic-estimation theoretic result, called the I-MMSE [1]-[2] to show that such tool generalizes time-frequency signal processing tools urgent for the analysis of non-stationary non-Gaussian signals.**

## I. INTRODUCTION

Stationary non-Gaussian signals are usually analyzed using tools from signal processing and statistics. Skewness and Kurtosis [3] are used to analyze the signal in the time domain. However, Fourier transform (FT) is used for non-stationary signals, i.e. signals whose spectral description is time dependent. On the other hand, the limitations of FT, enforces the use of time-frequency representation for two-dimensional signal processing analysis. The short time frequency transforms (STFT) [4] and the WD [5] aims to describe time dependent spectral properties, while the Wavelet Transform (WT) [6] aim at extracting localized contributions of signals. The STFT considers quasi-stationary signals and analyzes the signal by taking its FT. The time-frequency representation aims at overcoming the time versus frequency resolution trade-off. The WD[ǂ] of a signal $x(t)$ is given by:

$$W_x(t,f) = \int_{-\infty}^{\infty} x\left(t+\frac{\tau}{2}\right) x^*\left(t-\frac{\tau}{2}\right) e^{-j2\pi f\tau} d\tau \tag{1}$$

where $x^*(t)$ is the complex conjugate of $x(t)$

The amount of information loss due to the contamination of the signal due to noise or the due to sub-Nyquist sampling for continuous time stationary Gaussian [7] and non-Gaussian stationary signals has been characterized in [1],[2]. Yet the time-frequency signal processing tools stay weak to characterize non-stationary non-Gaussian signals. In this paper, we capitalize on information theoretic-estimation theoretic tools to show that such tools for stationary signals in the time-domain generalize time-frequency single processing tools. We characterize the amount of information loss as a function with respect to the minimum mean squared error as a function of the signal and the sampling process. Additionally, we show other terms that characterize the cross correlations in the process in the presence of multiple signals or the noise, all with respect to time and frequency.

ǂWe present the results with respect to WD $W_x(t,f)$ for ease of exploitation and for its direct relation to the power spectral density $S_x(f)$.

## II. DEFINITIONS

The non-linear estimator of a communication system is,

$$E[x(t)|y(t)] = \int_x x(t)p(x(t)|y(t)) \tag{2}$$

what represents a non-stationary process. Then,

$$\begin{aligned} E[(E[x(t)|y(t)])^2] &= \int_t \int_x x^2 p^2(x|y)dx\, dt \\ &= \int_t E_X[x^2 p^2(x|y)]dt \\ &= \int_{-\infty}^{\infty} S_{X|Y}(f)df \\ &= \int_{-\mathrm{f}_s/2}^{\mathrm{f}_s/2} S_{X|Y}(f)df \\ &= \sum_{-\infty}^{\infty} \int_{-\mathrm{f}_s/2}^{\mathrm{f}_s/2} W_{X|Y}(t,f)df \end{aligned} \tag{3}$$

(3) Where $S_{X|Y}(f)$ is the conditional power spectral density and $W_{X|Y}(t,f)$ is the conditional WD or the time-frequency counterpart. This term implies the amount of information loss due to sampling that persists after the reconstruction or estimation process.

The minimum mean squared error of the process in the time, frequency and time-frequency is given as,

$$mmse(t) = \int_{-\infty}^{\infty} E_X[(x(t) - E[x(t)|y(t)])^2]dt \tag{4}$$

$$mmse(f) = \int_{-\infty}^{\infty} S_X(f)df - \int_{-\mathrm{f}_s/2}^{\mathrm{f}_s/2} S_{X|Y}(f)df \tag{5}$$

$$\begin{aligned} mmse(t,f) = &\int_{-\infty}^{\infty}\int_{-\infty}^{\infty} W_X(t,f)dtdf \\ &- \sum_{-\infty}^{\infty} \int_{-\mathrm{f}_s/2}^{\mathrm{f}_s/2} W_{X|Y}(t,f)df \end{aligned} \tag{6}$$

where $W_X(t,f)$ is as in (1), and the conditional part [7] is,

$$\iint_{-\infty}^{\infty} W_{X|Y}(t,f)\, dtdf = \int_{-\mathrm{f}_s/2}^{\mathrm{f}_s/2} \frac{\sum_{k=-\infty}^{\infty} {S_X}^2(f-\mathrm{f}_s k,t)|H(f-\mathrm{f}_s k)|^2}{\sum_{k=-\infty}^{\infty} S_Y(f-\mathrm{f}_s k,t)|H(f-\mathrm{f}_s k)|^2} df$$

where $H(f - \mathrm{f}_s k)$ is the sampling filter.

## III. THE TIME-FREQUENCY I-MMSE

The following section presents a generalization tool for all time-frequency signal processing tool, which encapsulates all the terms of the contamination by noise, the reconstruction, the estimation, and the losses due to correlations of the signal with mixtures of other signals in the non-stationary non-Gaussian signal world. The I-MMSE in the time-frequency domain is a generalization to the ones presented in [1]-[2], and other presentations like ones in [8]-[13].

The connection between the mutual information and the minimum mean squared error defines the changes of the mutual information with respect to random parameters and this change characterizes the losses in the information encountered in the transmission, noise contamination and interference or mixtures of correlated and uncorrelated signals. This I-MMSE relation is defined with respect to the snr as follows:

$$\frac{dI(snr)}{dsnr} = mmse(snr) + \psi(snr) \qquad (7)$$

This can be re-written in the time-frequency domains as,

$$\frac{dI_{x;y,snr}(t,f)}{dsnr} = mmse_{x|y}(t,f) + \psi_{x|y}(t,f) \qquad (8)$$

Capitalizing on this fundamental result and the analysis in the previous section and considering a case of two signals mixtures $y = x_1 + x_2 + n$, we re-write this result in the time-frequency domains as follows,

$$\frac{dI_{x_1,x_2;y,snr}(t,f)}{dsnr} = mmse_{x_1|y}(t,f) + mmse_{x_2|y}(t,f) + \psi_{x_1x_2|y}(t,f) + \psi_{x_2x_1|y}(t,f) \qquad (9)$$

$$\begin{aligned}
\frac{dI_{x_1,x_2;y,snr}(t,f)}{dsnr} &= \int_{-\infty}^{\infty}\int_{-\infty}^{\infty} W_{x_1}(t,f)dtdf \\
&- \int_{-\infty}^{\infty}\int_{-f_s/2}^{f_s/2} \frac{W_{x_1}(t,f)}{W_y(t,f)}dtdf \\
&- \int_{-\infty}^{\infty}\int_{-\frac{f_s}{2}}^{\frac{f_s}{2}} \frac{W_{x_1x_2}(t,f)}{W_y(t,f)}dtdf \\
&- \int_{-\infty}^{\infty}\int_{-f_s/2}^{f_s/2} \frac{W_{x_1x_2}(t,f)W_{x_1}(t,f)}{W_y(t,f)}dtdf \\
&- \int_{-\infty}^{\infty}\int_{-f_s/2}^{f_s/2} \frac{W_{x_1}(t,f)W_{x_1^*x_2^*}(t,f)}{W_y(t,f)}dtdf \\
&+ \int_{-\infty}^{\infty}\int_{-\infty}^{\infty} W_{x_2}(t,f)dtdf \\
&- \int_{-\infty}^{\infty}\int_{-f_s/2}^{f_s/2} \frac{W_{x_2}(t,f)}{W_y(t,f)}dtdf \\
&- \int_{-\infty}^{\infty}\int_{-\frac{f_s}{2}}^{\frac{f_s}{2}} \frac{W_{x_2x_1}(t,f)}{W_y(t,f)}dtdf \\
&- \int_{-\infty}^{\infty}\int_{-f_s/2}^{f_s/2} \frac{W_{x_2x_1}(t,f)W_{x_2}(t,f)}{W_y(t,f)}dtdf \\
&- \int_{-\infty}^{\infty}\int_{-f_s/2}^{f_s/2} \frac{W_{x_2}(t,f)W_{x_2^*x_1^*}(t,f)}{W_y(t,f)}dtdf \\
&- \int_{-\infty}^{\infty}\int_{-f_s/2}^{f_s/2} \frac{W_{x_1}(t,f)W_{x_2}(t,f)}{W_y(t,f)}dtdf \\
&- \int_{-\infty}^{\infty}\int_{-f_s/2}^{f_s/2} \frac{W_{x_1}(t,f)W_{x_2^*x_1^*}(t,f)}{W_y(t,f)}dtdf \\
&- \int_{-\infty}^{\infty}\int_{-f_s/2}^{f_s/2} \frac{W_{x_1x_2}(t,f)W_{x_2^*x_1^*}(t,f)}{W_y(t,f)}dtdf \\
&- \int_{-\infty}^{\infty}\int_{-f_s/2}^{f_s/2} \frac{W_{x_1x_2}(t,f)W_{x_2}(t,f)}{W_y(t,f)}dtdf \\
&- \int_{-\infty}^{\infty}\int_{-f_s/2}^{f_s/2} \frac{W_{x_2}(t,f)W_{x_1}(t,f)}{W_y(t,f)}dtdf \\
&- \int_{-\infty}^{\infty}\int_{-f_s/2}^{f_s/2} \frac{W_{x_2}(t,f)W_{x_1x_2}(t,f)}{W_y(t,f)}dtdf \\
&- \int_{-\infty}^{\infty}\int_{-f_s/2}^{f_s/2} \frac{W_{x_2x_1}(t,f)W_{x_2^*x_1^*}(t,f)}{W_y(t,f)}dtdf \\
&- \int_{-\infty}^{\infty}\int_{-f_s/2}^{f_s/2} \frac{W_{x_2x_1}(t,f)W_{x_1}(t,f)}{W_y(t,f)}dtdf \qquad (10)
\end{aligned}$$

This result generalizes the I-MMSE identity [1]-[2] for single and multiple inputs –output systems to an I-MMSE in the time-frequency domains. This result generalizes also the WD [5] distribution by adding terms that explain explicitly the loss of information that exist in a non-Gaussian non stationary mixture i.e. the effect of sampling and estimation and the effect of interference or contamination due to multiple inputs, which conform with the definition of the mutual information or the Kullback Leibler distance [14] that expresses the distance between two distributions. The WD of a mixture is the sum of the WD of the two inputs minus the real of their cross correlation. However, this expression above provides all the positive and negative terms when the inputs are correlated and complex with the presence of the cross correlation between the inputs estimates. If the inputs are real,

$$\begin{aligned}
&\frac{dI_{x_1,x_2;y,snr}(t,f)}{dsnr} \\
&= \int_{-\infty}^{\infty}\int_{-\infty}^{\infty} W_{x_1}(t,f)dtd \\
&- \int_{-\infty}^{\infty}\int_{-f_s/2}^{f_s/2} \frac{W_{x_1}(t,f)}{W_y(t,f)}dtdf \\
&+ \int_{-\infty}^{\infty}\int_{-\infty}^{\infty} W_{x_2}(t,f)dtdf - \int_{-\infty}^{\infty}\int_{\frac{-f_s}{2}}^{\frac{f_s}{2}} \frac{W_{x_2}(t,f)}{W_y(t,f)}dtdf \\
&- 2Re\left\{\int_{-\infty}^{\infty}\int_{-\frac{f_s}{2}}^{\frac{f_s}{2}} \frac{W_{x_1x_2}(t,f)}{W_y(t,f)}dtdf\right\} \\
&- 4Re\left\{\int_{-\infty}^{\infty}\int_{-f_s/2}^{f_s/2} \frac{W_{x_1x_2}(t,f)W_{x_1}(t,f)}{W_y(t,f)}dtdf\right\} \\
&- 4Re\left\{\int_{-\infty}^{\infty}\int_{-f_s/2}^{f_s/2} \frac{W_{x_1x_2}(t,f)W_{x_2}(t,f)}{W_y(t,f)}dtdf\right\} \\
&- 2Re\left\{\int_{-\infty}^{\infty}\int_{-\frac{f_s}{2}}^{\frac{f_s}{2}} \frac{W_{x_1}(t,f)W_{x_2}(t,f)}{W_y(t,f)}dtdf\right\} \\
&-2Re\left\{\int_{-\infty}^{\infty}\int_{-f_s/2}^{f_s/2} \frac{\left(W_{x_2x_1}(t,f)\right)^2}{W_y(t,f)}dtdf\right\} \qquad (11)
\end{aligned}$$

Under the independence assumption, i.e. when there is no correlation, the rest of the terms are zero. The equation reduces to the 1st, 2nd, 6th, 7th, 11th, and 15th as follows,

$$
\begin{aligned}
\frac{dI_{x_1,x_2;y,snr}(t,f)}{dsnr} = &\int_{-\infty}^{\infty}\int_{-\infty}^{\infty} W_{x_1}(t,f)\,dtdf \\
&- \int_{-\infty}^{\infty}\int_{-\mathrm{f}_s/2}^{\mathrm{f}_s/2} \frac{W_{x_1}(t,f)}{W_y(t,f)}\,dtdf \\
&+ \int_{-\infty}^{\infty}\int_{-\infty}^{\infty} W_{x_2}(t,f)\,dtdf \\
&- \int_{-\infty}^{\infty}\int_{-\mathrm{f}_s/2}^{\mathrm{f}_s/2} \frac{W_{x_2}(t,f)}{W_y(t,f)}\,dtdf \\
&- \int_{-\infty}^{\infty}\int_{-\frac{\mathrm{f}_s}{2}}^{\frac{\mathrm{f}_s}{2}} \frac{W_{x_1}(t,f)W_{x_2}(t,f)}{W_y(t,f)}\,dtdf \\
&- \int_{-\infty}^{\infty}\int_{-\frac{\mathrm{f}_s}{2}}^{\frac{\mathrm{f}_s}{2}} \frac{W_{x_2}(t,f)W_{x_1}(t,f)}{W_y(t,f)}\,dtdf \qquad (12)
\end{aligned}
$$

Notice also that the WD is real, but the inputs might not be real, to which we did not simplify some terms, to reflect the complex part, especially that unlike the normal distribution the WD could be negative and this in effect makes interfering terms sometimes of benefit.

## CONCLUSIONS

In this paper, we derive a time-frequency I-MMSE version. This helps generalize time-frequency signal processing tools like the WD by considering the estimation and reconstruction of non-stationary non-Gaussian signals contaminated by similar signals. The result applies to the Gaussian and stationary inputs as a special case.